# Visual Function Profiles via Multi-Path Aggregation Reveal Neuron-Level Responses in the Drosophila Brain

Jiangping Xie [A], Ruohan Ren [B, C], Xiao Zhou [A], Ao Zheng [D, E], Jiasong Zhu [E, F], Wenyu Jiang [E, F, *], Ziran Zhao [A, *]

[A] Department of Engineering Physics, Tsinghua University, Beijing, 100084, China

[B] Computational and Systems Biology Program, Sloan Kettering Institute, Memorial Sloan Kettering Cancer Center, New York City, NY, USA

[C] Tri-Institutional Training Program in Computational Biology and Medicine, Weill Cornell Medicine, New York City, NY, USA

[D] Department of Civil and Environmental Engineering, The Hong Kong Polytechnic University, Hong Kong, China

[E] College of Civil and Transportation Engineering, Shenzhen University, Shenzhen, China

[F] MNR Key Laboratory for Geo-Environmental Monitoring of Great Bay Area, Shenzhen, 518060, China

[*] Corresponding author

**Abstract：**

Accurately predicting individual neurons' responses and spatial functional properties in complex visual tasks remains a key challenge in understanding neural computation. Existing whole-brain connectome models[1,2] of *Drosophila* often rely on parameter assumptions or deep learning approaches, yet remain limited in their ability to reliably predict dynamic neuronal responses. We introduce a Multi-Path Aggregation (MPA) framework, based on neural network steady-state theory, to build a whole-brain Visual Function Profiles (VFP) of Drosophila neurons and predict their responses under diverse visual tasks. Unlike conventional methods relying on redundant parameters, MPA combines visual input features with the whole-brain connectome topology. It uses adjacency matrix powers and finite-path optimization to efficiently predict neuronal function, including ON/OFF polarity, direction selectivity, and responses to complex visual stimuli. Our model achieves a Pearson correlation of 0.84±0.12 for ON/OFF responses, outperforming existing methods (0.33±0.59[1]), and accurately captures neuron functional properties, including luminance and direction preferences, while allowing single-neuron or population-level blockade simulations. Replacing CNN modules with VFP-derived Lobula Columnar (LC) population responses in a *Drosophila* simulation[3] enables successful navigation and obstacle avoidance, demonstrating the model's effectiveness in guiding embodied behavior. This study establishes a "connectome-functional profile-behavior" framework, offering a whole-brain quantitative tool to study *Drosophila* visual computation and a neuron-level guide for brain-inspired intelligence.

## 1 Main

In contemporary neuroscience, researchers can map the structural connectivity of neural circuits, such as the *Drosophila* brain connectome[4-11]. However, it remains challenging to simultaneously capture other biological properties underlying connectivity. To investigate how visual neural processes can inform artificial intelligence and to uncover the functional mechanisms of biological neural circuits, previous studies have constructed whole-brain models of *Drosophila* based on neuronal structural connectivity[1,2], aiming to partially reconstruct neural activity through modeling. These studies indicate that deep learning models leveraging circuit connectivity and task-specific knowledge can predict the contributions of individual neurons in computations, nevertheless discrepancies between model predictions and experimental data often persist. Furthermore, such approaches typically rely on several parametric assumptions, which may limit model interpretability and applicability.

From a reductionist perspective, we ask whether neural activity can be partially reconstructed using only connectome and neural response data. Recent studies in *Drosophila*, spanning from synapses to the compound eye[12-16], suggest that incorporating synaptic density and neuronal projections to specific compound eye regions, in addition to connectome data, can more accurately capture the spatial organization of visual information processing.

Motivated by these findings, we focus on the *Drosophila* visual system and apply network topology analysis methods[17-19] within a connectomics framework. Using a whole-brain connectivity matrix of nearly 140,000 neurons[20-22] and relevant graph optimization techniques, we propose a Multi-Path Aggregation (MPA) method based on a steady-state neural system. This approach generates visual function profiles (VFP) for individual neurons across the *Drosophila* brain, enabling the prediction of neuronal response patterns to diverse visual stimuli.

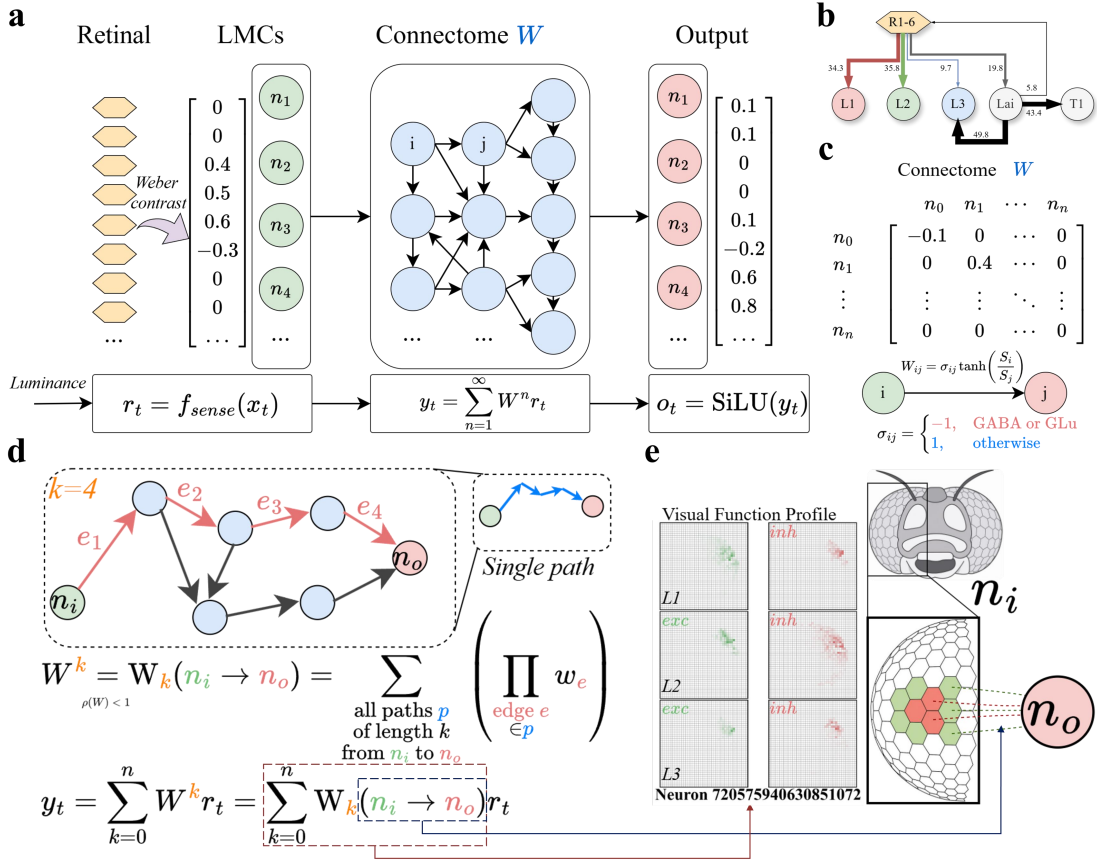

**Figure 1. VFP model.** (a) Model schematic: an image luminance sequence is processed by LMCs and combined with the whole-brain connectivity matrix to compute matrix powers, producing the network steady state. Low-pass filtering and nonlinear transformation are then applied to align the output neural activity with biological observations. (b) Connections between LMCs and photoreceptors R1–R6; arrow values indicate the proportion of synapses derived from connectome data. (c) Illustration of the whole-brain neuronal connectivity matrix of *Drosophila*, with values obtained from FLYWIRE data. (d) Schematic of the MPA method: a finite-path optimization is applied to the large-scale connectivity matrix to compute the association strength between LMCs and other neurons. (e) Contributions of the LMCs to output neurons obtained via the optimization in (d), mapped onto the *Drosophila* compound eye cylindrical coordinates (size 41×41) to generate a six-dimensional VFP.

Using adjacency matrix powers[23] and finite-path optimization, we developed the MPA method to generate VFP for all neurons in the *Drosophila* brain. Its main advantage is that it directly leverages whole-brain connectome data, integrating visual inputs with synaptic connectivity weights to model the structure of neuronal response patterns across the brain.

As shown below, combining these VFP enables the model to efficiently infer steady-state activity in large-scale networks and accurately predict ON/OFF responses[24-27], direction selectivity[27-29], and neural responses under diverse visual stimuli[30,31]. The model is modular and scalable, allowing the inclusion of silenced neurons[27,32-35] or the replacement of parts of artificial neural networks[3].

Quantitative analyses demonstrate that our model outperforms existing approaches in both accuracy and stability. The results indicate that VFP-based prediction models can effectively simulate neuronal activity during visual processing in *Drosophila*, providing a quantitative computational tool for understanding the functional organization of the fly visual system. Finally, we provide a web-based tool for dynamic visualization of neuronal responses under varying visual inputs (for details, see Supplementary Materials).

## 1.1 ON/OFF Function Response

In the early stages of visual information processing, neural pathways generally exhibit selective responses to contrast changes, with ON pathways specialized for detecting contrast increments and OFF pathways encoding contrast decrements[25,36-38]. This functional segregation is particularly pronounced in the primary visual layers, where response properties show differential sensitivity to luminance and contrast parameters.

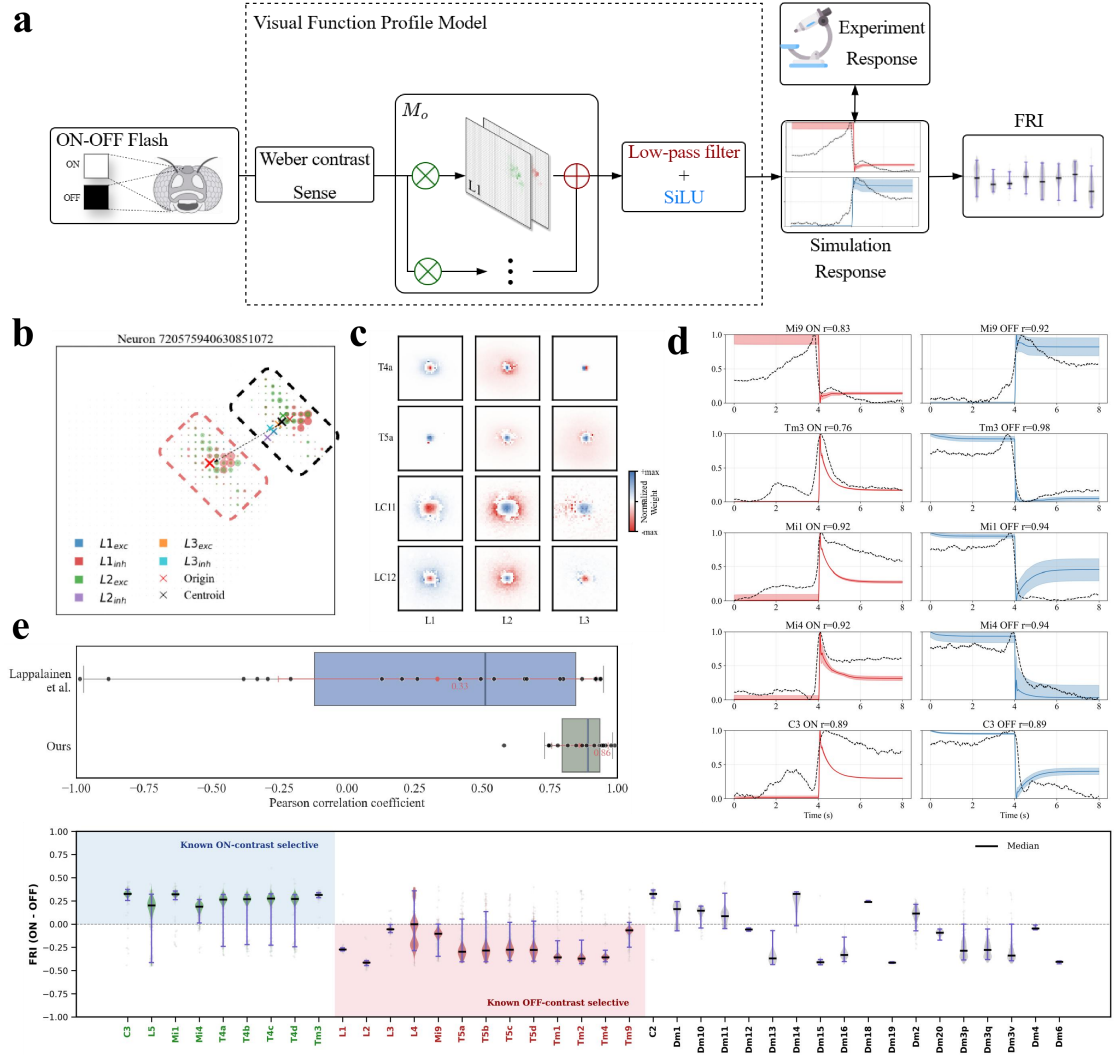

**Figure 2. Validation of the VFP model on known ON/OFF neurons.** (a) Workflow of the model validation experiment. (b) Illustration of the overall centroid shift in the individual-neuron VFP. (c) Average distribution of the shifted VFP for selected neuron categories. (d) Comparison between model outputs and experimentally measured ON/OFF response curves. (e) Pearson correlation coefficients between model outputs and 22 experimental response datasets, compared with the baseline model[1]; additional experimental comparison curves are shown in Appendix Figure 2(a). (f) ON- and OFF-contrast selectivity indices (FRI) for selected neuron categories, with blue indicating neurons known to respond to ON contrast and red indicating neurons known to respond to OFF contrast.

Based on this, we implemented the model validation workflow shown in Figure 2(a). The model receives ON/OFF stimulus sequences as input, applies Weber contrast processing and nonlinear transformation at the Lamina Monopolar Cells (LMCs)[39,40], and then performs a dot-product summation with the VFP. Neuronal responses are subsequently obtained via low-pass filtering and the SiLU nonlinear function[41], and FRI distributions are plotted for different neuron categories. Finally, these results are compared with experimental data.

We observed that different neurons are sensitive to distinct regions of the visual field. To spatially align the VFP of individual neurons, we first computed the overall centroids of their excitatory and inhibitory VFP. These centroids reflect the positions on the 2D grid associated with each neuron of the *Drosophila* compound eye, representing the spatial distribution of LMCs' contributions to the neuron.

Next, for each visual neuron type, we averaged the shifted VFP across all neurons to visualize the spatial distribution of inputs from LMCs, as shown in Figure 2(c). This averaged profile represents the receptive field distribution of a given neuron type on the retina. Spatial differences in visual input across neurons of the same type reveal functional diversity and heterogeneity in their responses. Compared to synapse-mapping-based methods[12,15], our approach more directly captures the weighted spatial distribution of visual inputs, providing a physiologically meaningful description of neuronal input features at the functional level.

To quantify single-neuron preferences for "ON-OFF" stimuli, we employed the FRI metric computed from simulated neuron-layer responses[1]. ON and OFF stimulus sequences were applied to the simulated network to measure peak responses, assessing each neuron's stimulus-type selectivity. To ensure comparability across conditions, ON-OFF timings were aligned. We compared the previous optimal model[1] with our model by presenting identical stimuli and computing Pearson correlation coefficients between their response curves. Figures 2(d)-(f) show that our model not only achieves higher accuracy in predicting temporal ON/OFF responses but also outperforms existing models in capturing neuron-specific ON/OFF selectivity.

It is important to note that the model assesses neurons based on the actual function of each individual neuron within the whole-brain network, rather than relying on population statistics or parameter fitting. Although population-level visualizations are used for presentation purposes, these categories are solely intended to facilitate interpretation and comparison of neuronal response properties. The underlying analysis preserves the independent functional assessment of every neuron in the brain.

## 1.2 Direction Function Response

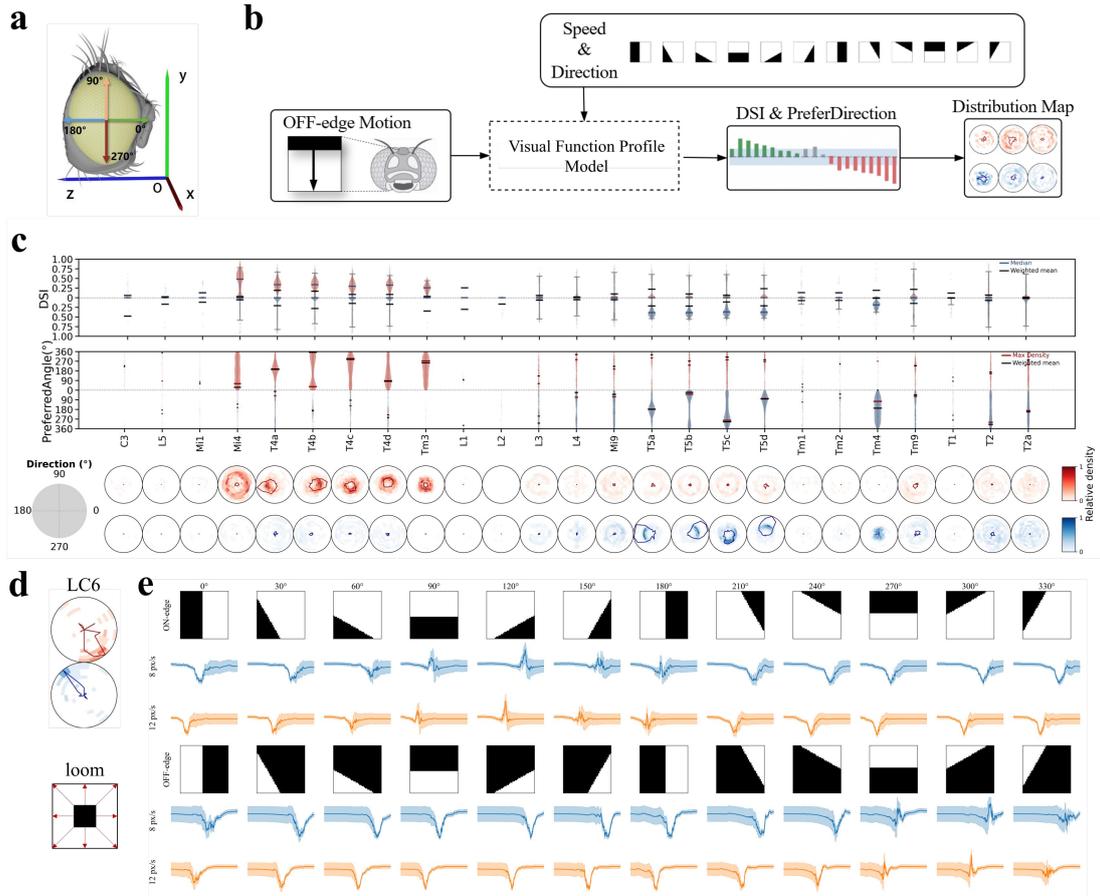

**Figure 3. Validation of the model on neurons with known direction selectivity.** (a) Illustration of the stimulus angles used in the model and their geometric relationship with the *Drosophila* compound eye. (b) Workflow diagram showing the steps for computing neuronal responses and analyzing direction selectivity. (c) Distribution of direction selectivity indices (DSI) and preferred angles for selected neurons, along with corresponding response heatmaps and tuning curves. (d) Tuning curves of the LC6 and their corresponding preferred stimuli in biological experiments. (e) Response curves of LC6 (median ± SEM) to ON/OFF edge stimuli at different directions.

We employed a Direction Selectivity Index (DSI) calculation method based on simulated neuronal responses[42] to quantify the directional preference of individual neurons in the *Drosophila* visual system. Specifically, for each neuron, we designed ON and OFF edge stimuli moving in multiple directions at different speeds and fed these stimuli into the model. Neuronal time-series responses were then simulated, and the DSI for each neuron was computed to assess its selectivity for specific motion directions.

Figure 3(c) shows the distribution of DSI for selected neuron classes. As seen, within the model, both T4 and T5 neuron subclasses exhibit distinct directional preferences at the level of individual neurons. These preferences are highly consistent with previous experimental results[27-29,35], indicating that our model accurately captures direction selectivity at the single-neuron level in the *Drosophila* visual system.

Notably, the distributions in the figure represent the preferences of individual neurons within each subclass, reflecting the variability in DSI across neurons of the same type. Analyzing the DSI distribution at the single-neuron level, rather than using only the mean preferred direction, preserves the heterogeneity of direction selectivity within a subclass, which is advantageous for subsequent functional analyses in combination with connectome data. This approach allows for a more precise investigation of functional differences within neuron subclasses and their contributions to overall motion encoding. Similarly, for the Mi4 neuron class shown in Figure 3(c), although the class as a whole does not display strong direction selectivity, individual neurons within the subclass may still exhibit specific directional preferences.

In addition to the T4 and T5 subclasses, we found that LC6 neurons may also exhibit direction selectivity. LC6 is known to respond preferentially to looming dark-object stimuli[43,44]. In our simulations, we observed that LC6 neurons respond significantly more strongly to dark-object stimuli expanding from the ventral side than to stimuli of other directions or luminance. This observation suggests that LC6 may have functional selectivity for processing visual threat signals from specific directions. Notably, previous studies have shown that LC6 activation can trigger backward jumping behavior in *Drosophila*[45], indicating that its direction selectivity may reflect a behavioral function related to avoiding predators approaching from below.

### 1.3    Model Response to Neuronal Blocking

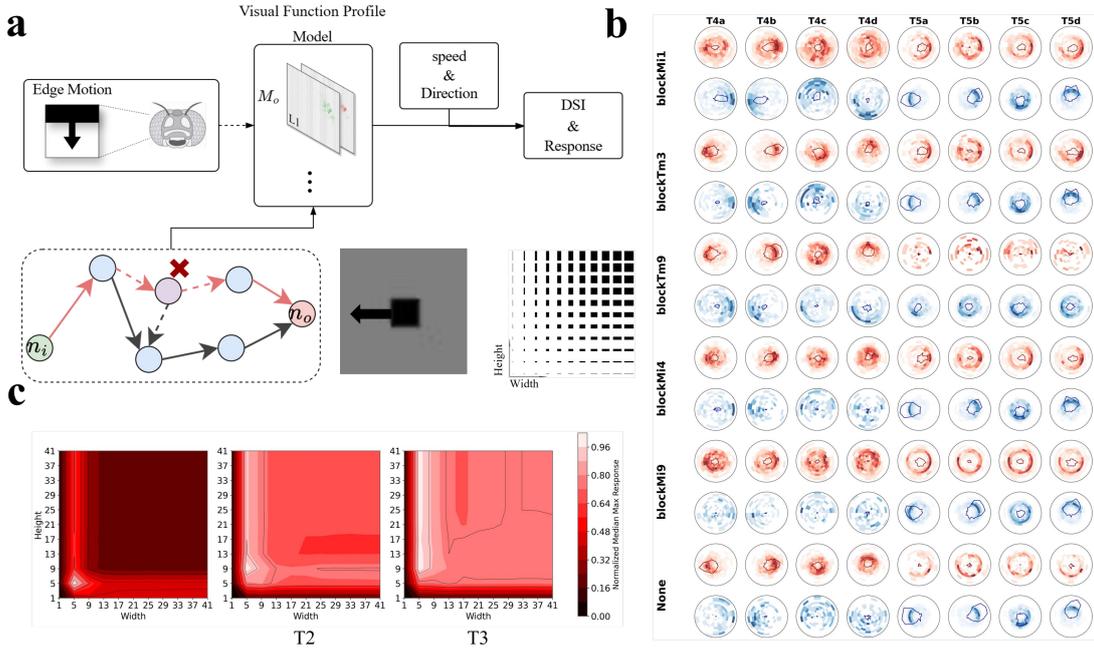

**Figure 4. Model responses under neuronal blocking.** (a) Schematic of the blocking model and responses of LC11 to stimuli of different sizes. (b) Response heatmaps and tuning curves of T4/T5 neurons after blocking upstream neurons contributing to their direction selectivity. (c) Responses of LC11 to moving dark objects of different sizes, and corresponding heatmaps after blocking T2 and T3 neurons.

Our model surpasses previous studies in terms of coverage depth within neural circuits. It can predict responses of neurons across the whole brain and simulate functional blocking experiments for individual neurons or specific neural populations. To implement neuronal blocking, the targeted neuron or neuron class is removed during the computation of the whole-brain connectivity matrix. The corresponding VFP under blocking is then recalculated and used as input in the model to compute visual responses.

Figure 4(b) shows that blocking upstream neuron classes of the T4/T5 subclasses affects their direction selectivity to varying degrees. We found that Mi1, Mi4, and Mi9 cells are absolutely essential for the normal operation of the ON pathway. As shown in the figure, blocking Mi1, Mi4, or Mi9 severely reduces the direction selectivity of T4 neurons, shrinking the tuning curves for all T4 subclasses, which indicates a decrease in the number of direction-selective neurons and a reduction in DSI values, consistent with experimental observations[35,46]. Although blocking Tm3 narrows the tuning curves of T4 subclasses, it has minimal impact on overall direction selectivity.

Previous experiments have shown that blocking Tm3 mainly impairs sensitivity to high edge speeds[47]. Finally, blocking Tm9 in the model severely affects T5 direction selectivity, consistent with biological experiments[48,49].

LC11 neurons are widely recognized as being responsible for detecting small moving objects in the *Drosophila* visual system. Previous studies have shown that blocking their main upstream inputs, T2 and T3 neurons, can partially impair LC11's target detection ability[34]. In our simulations, we observed that blocking T3 outputs significantly reduces LC11's response discrimination across objects of different sizes, particularly diminishing the difference between small and large objects. However, LC11 still maintains a relative preference for the smallest objects. Similarly, blocking T2 slightly reduces LC11's response to small objects, but the overall response pattern remains similar to the normal state. These results indicate that T2 and T3 cooperate in LC11's small-object detection, with T3 likely playing a more critical role in size selectivity modulation.

For LC neurons, some are known to exhibit size specificity to different stimuli[30,31]. We calculated the responses of numerous LC neurons to objects of various sizes and obtained median responses to visually illustrate each neuron type's size specificity. Responses of other LC neurons to different object sizes are shown in Appendix Figure 2(b).

### 1.4    LC Response–Based Neural Network Model

The design of artificial neural networks is inspired by biological neural networks[50-52]. To further investigate whether our model can faithfully capture neural signals and enhance the performance of artificial neural networks, we embedded the visual pathway model constructed in this study into the *Drosophila* whole-body simulation platform[3], replacing the convolutional neural network (CNN) previously used for high-level visual navigation in a trench task. This replacement allows us to study the role of the *Drosophila* visual system in complex scene perception and motor control in a biologically inspired manner.

LC populations serve as output units for higher-order visual processing in *Drosophila*, primarily located in the Lobula[5,53]. They convey integrated information from early visual pathways (e.g., L1–L3 → Medulla → Lobula) to downstream targets. Some LC neurons, such as LC10 and LC11, are highly sensitive to motion direction or object movement and can modulate turning or flight speed[32,45,54,55]. Acting as a critical bridge between visual input and motor control, LC neurons integrate directional selectivity, ON/OFF selectivity, and optic flow to drive downstream motor neurons, thereby enabling real-time flight, turning, and obstacle avoidance.

Within this framework, we first designed an encoding module based on LMCs responses to convert RGB images into simulated LMCs responses in the model, following the same computational method as the $f_{sence}$ module. This involves computing local contrast changes in the receptive fields to obtain simulated LMCs neuronal activity.

Since our model corresponds only to the right-eye visual pathway, we used the left-right mirror pairing information from the FLYWIRE dataset to match corresponding LC neurons between the two eyes. Through this mirroring, we can get the VFP of the left eye to generate left-side LC responses. For each LC type, responses of all neurons in that type are linearly projected into an 8-dimensional vector, which we term a subclass controller. All LC types are then combined through linear weighting and integrated with other body parameters into a higher-level controller[56], which extracts higher-order spatial and directional information and provides input to the low-level controller in the simulation platform for driving behavioral outputs such as turning and obstacle avoidance. Following[57], we selected 16 representative LC types as candidate types. After reinforcement learning training, as shown in Figures 5(b)-(c), we found that the simple linear LC network can effectively replace the previous CNN network.

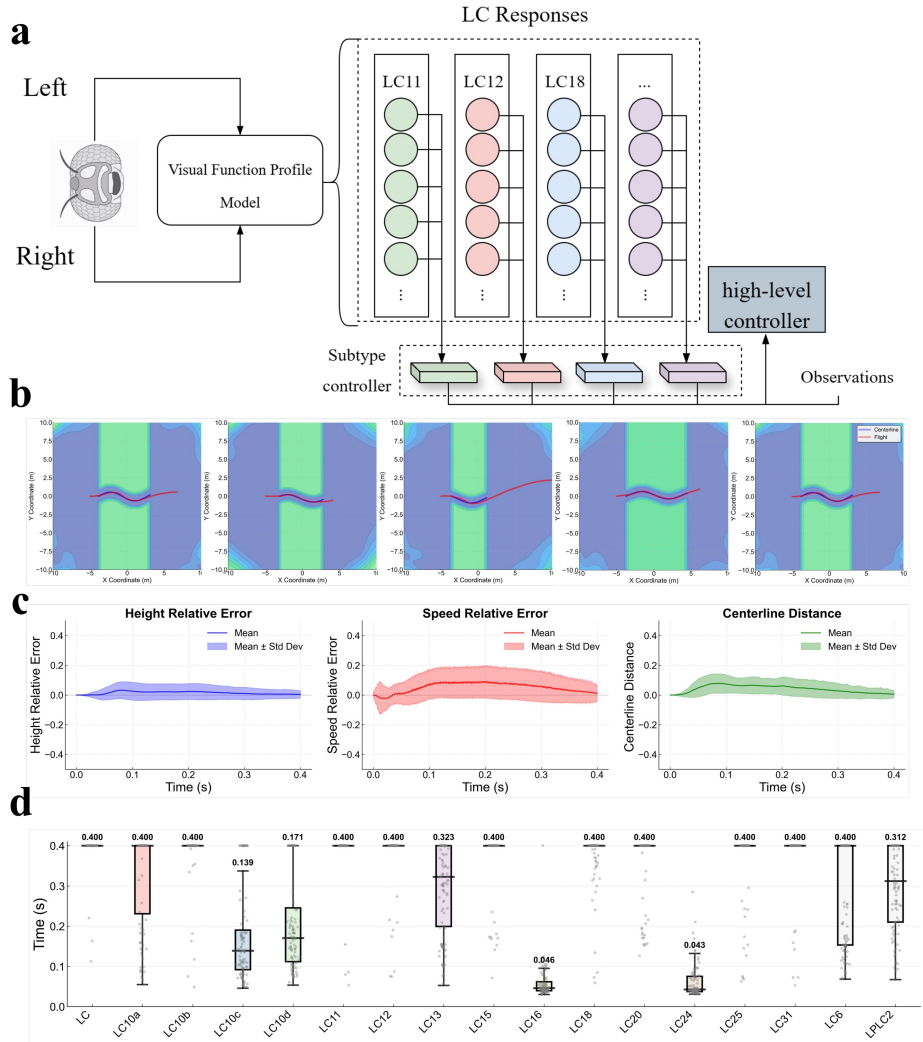

**Figure 5. LC-Based Model.** (a) Schematic of the LC model. (b) Top-down view of flight trajectories in five test scenarios after reinforcement learning training. (c) Relative errors in altitude and speed (percentage) and distance errors (in meters) across 1,000 random scenarios. (d) Effect on fly survival time of zeroing out the VFP of certain LC subclass neurons in the model, relative to the unmanipulated LC reference.

Additionally, after network training, we performed a zeroing operation on the VFP of certain candidate LC types in the model, effectively deactivating the corresponding 8-dimensional subclass controllers to simulate the removal or loss of these connections. By selectively blocking different subclass controllers one by one, we can observe changes in network performance, particularly deviations in behavior and decreases in task performance during navigation and obstacle avoidance. Figure 5(d) shows that blocking LC6, LC10a, LC10c, LC10d, LC13, LC16, LC24, and LPLC2 leads to significant navigation path deviations and increased obstacle collision rates.

## 2    Discussion

In this study, we propose a connectome-based VFP for individual neurons, aiming to simulate and predict neuronal responses to specific visual stimuli in the *Drosophila* visual system. The model fully leverages the structural connectivity information of neurons in the compound eye, encoding functional pathways between neurons as a weighted topological network. This allows the model to capture not only the response properties of single neurons but also the collective behavior of neuronal populations during visual information processing. By introducing the MPA method, the model can represent the strength of influence and information flow between neurons, enabling high-precision predictions of neuronal activity under complex visual stimuli, and providing a quantitative computational tool for understanding the functional organization of the *Drosophila* visual system.

In simulation experiments, we validated the model's predictions of ON-OFF response characteristics and direction selectivity of visual neurons. The results show that the proposed VFP outperforms previous models in capturing neuronal sensitivity to luminance changes and motion direction. This improvement indicates that incorporating connection weights and topological path information allows the model to more accurately reflect functional dependencies within the neural network, thereby enhancing prediction of individual neuronal responses. Furthermore, these results suggest that direction selectivity and luminance responses in the *Drosophila* visual system may not only depend on local properties of individual neurons but are also strongly regulated by their network position and connectivity with LMCs. By explicitly modeling these path strengths, our model captures, to some extent, such network-level regulatory mechanisms, providing new insights into population-level neural dynamics in compound eye visual processing.

Moreover, the improved performance of the model provides a reference for future neuroscience research and computational neural modeling. By integrating structural connectivity information with functional responses, the model serves as an effective tool for exploring the principles underlying visual processing of complex stimuli. This approach also highlights the potential of topological and weighted information for simulating other biological neural systems, offering a generalizable framework for cross-species neural network modeling.

Using the *Drosophila* whole-body simulation platform, we embedded our visual pathway model into the framework to biologically and inspiredly replace the original CNN module's high-level visual processing, thereby assessing the role of the *Drosophila* visual system in complex terrain navigation. By encoding LMCs responses, pairing left and right eyes, and integrating type-specific features, early visual signals are transformed into high-level spatial features and fed into a low-level controller to drive behavior generation. This integrative strategy not only validates the utility of VFP in motor control tasks but also demonstrates the importance of biologically grounded visual processing in behavior modulation.

It should be noted that the MPA-based VFP model proposed in this study remains constrained by a reductionist framework based on connectome data. To avoid introducing additional parameters, we employed a purely mathematical framework combined with nonlinear operations to simulate neuronal responses, without using spiking neuron models or deep learning-derived parameters. Importantly, the model predicts single-neuron activity, whereas experimental measurements typically rely on calcium signals reflecting population activity. Due to the complex conversion from voltage to calcium signals[58] and the modulation of population dynamics by spatial and topological factors, the model may not precisely capture ensemble calcium dynamics. Although the model offers advantages in predicting visual responses, its applicability to complex cognitive tasks and higher-order behaviors remains unvalidated. Model performance is also limited by data completeness and annotation accuracy, affecting generalizability and interpretability. Future improvements could integrate population-level responses and spatial distribution to enhance prediction accuracy.

Several directions remain promising for future investigations. First, as the current network is primarily steady-state based, incorporating synaptic plasticity[59-61] could allow real-time dynamic adjustment of synapses to better capture temporal dependencies. Second, one could leverage activity patterns and dynamics of motor control neurons[62] to transform visual or other sensory inputs into concrete behavioral outputs. Finally, the model could be extended to multimodal contexts, especially olfaction[63-65], to study neuronal responses under combined stimuli and to explore the coupling between connection weights and functional responses.

## 3    Method

### 3.1    Visual Inputs from LMCs Contrast Responses

In the Weber contrast calculation, the target and background luminances are denoted as $I_{target}$ and $I_{background}$, respectively. In the model, the luminance of the current frame $x_t$, represents the target luminance, while the previous frame, $x_{t-1}$, represents the background luminance. The response of LMCs under different Weber contrasts is modeled as:

$$C_W(x_t) = \frac{I_{target} - I_{background}}{I_{background}} = \frac{x_t - x_{t-1}}{x_{t-1}}, f_{sence}(C_W) = \frac{aC_W}{1 + |bC_W|}$$

Here, the parameters $a, b$ were quantitatively fitted to electrophysiological data [39,40] to capture LMCs response patterns under different Weber contrast stimuli. The fitted curve follows a sigmoid shape, as shown in Appendix Figure 1.

The network response $r_t$ is then computed as:

$$r_t = f_{sence}(C_W(x_t))$$

where $x_t$ represents the external luminance change, and $r_t$ represents the internally generated network signal rather than the externally applied stimulus. This modeling approach provides a mathematically grounded representation of contrast processing in the early visual system, offering physiologically plausible inputs for subsequent layers and allowing LMCs responses to be derived directly from input images.

### 3.2 Multi-Path Aggregation

Based on the connectome data from the FlyWire dataset[4-11], we constructed a whole-brain connectivity graph of *Drosophila* comprising approximately 140,000 neurons. Mathematically, the connectome is formalized as a directed weighted graph, where neurons serve as nodes and synaptic connections as edges. By quantifying the synaptic strength between neurons, we defined an adjacency matrix $W$ of dimension approximately $140,000 \times 140,000$. Each matrix element $w_{ij}$ is represented by the normalized synapse count, defined as:

$$w_{ij} = \sigma_{i,j} \, tanh\left(\frac{n_{i,j}}{n_j}\right)$$

where $n_{i,j}$ is the number of synapses from neuron $i$ to neuron $j$, $n_j$ is the total input synapses to neuron $j$, and $\sigma_{i,j}$ follows the convention in[1,2], taking -1 for inhibitory (GABAergic or Glutamatergic) connections and 1 otherwise. This formulation ensures the boundedness of the connection weights ($0 \leq w_{ij} < 1$) and incorporates competitive normalization among input synapses, while normalizing the sum of positive and negative weights for each neuron for computational convenience. Due to the inherent sparsity of biological connectomes, $W$ is highly sparse.

Spectral analysis via power iteration yields a spectral radius $\rho(W) \approx 0.84 < 1$. According to Gelfand's formula[66], this spectral condition guarantees the convergence of the matrix power sequence $W^k (k \to \infty)$, and thus the convergence of the power series $\sum_{k=0}^{\infty} W^k$. This ensures that the network's response to any internal input $r_t$ does not diverge, maintaining model stability.

From a biological perspective, the successive powers $W^k$ and the cumulative series $\sum_{k=0}^{\infty} W^k$ carry clear neurodynamic interpretations. Specifically, $W^k$ captures indirect influence paths spanning $k$ synapses: $k = 1$ corresponds to direct synaptic input, while $k \geq 2$ captures multi-step signal propagation through intermediate neurons. As $k$ increases, $W^k$ represents modulation effects over longer chains, potentially including recurrent loops, feedback, lateral pathways, and cross-regional indirect regulation. Summing contributions across all orders ($\sum_{k=0}^{\infty} W^k$) can be interpreted as a full-path integration of the network over time, reflecting the long-term steady-state influence of each neuron on all others rather than merely instantaneous direct effects.

### 3.3 Spectrally Constrained Steady-State Network Response

In the model, the *Drosophila* whole-brain connectome is formalized as a normalized weighted adjacency matrix $W \in \mathbb{R}^{N \times N}$, and the network's response to an internal input $r_t \in R^N$ is defined as the steady-state output:

$$o_t = \sum_{k=0}^{\infty} W^k r_t = (I - W)^{-1} r_t$$

The convergence of the power series requires the spectral radius $\rho(W) < 1$. This definition is equivalent to the steady-state solution of a discrete-time linear dynamical system[67,68], $x_\infty = (I - W)^{-1} r_t$, ensuring that the network does not diverge under any input and thus maintains stable signal propagation. Biologically, the steady-state output $o_t$ reflects the balanced activity of the network under internal stimulation $r_t$, and the normalization of excitatory and inhibitory synaptic weights ensures the balance between excitation and inhibition[69], producing stable network responses.

### 3.4 Visual Function Profile (VFP)

For the steady-state solution, directly computing $(I - W)^{-1}$ for a sparse matrix of size $N \sim 140000$ is extremely memory-intensive, so optimization is necessary. Since the inputs considered in this work come from the LMCs, we propose the MPA method to simplify the analysis to calculating the contributions of LMCs to each neuron across the whole brain. According to the connectome data, the right-eye LMCs in Drosophila consist of approximately 800 columnar neurons (mapped to a $41 \times 41$ matrix in this model). On this basis, the contribution of signals flowing through LMCs can be accurately computed by quantifying all possible synaptic paths. For a given input neuron $n_i$ (located in LMCs), output neuron $n_o$, and path length $k$, the connection weight is defined as[20,21]:

$$W^k \approx W_k(n_i \to n_o) = \sum_{P_k} \prod_e w_e$$

Here, $P_k$ denotes the set of all paths of length $k$ from $n_i$ to $n_o$, and $w_e$ is the weight of a single synaptic connection $e$ along the path $P_k$.

Considering the physical decay in biological synaptic transmission, the influence of a signal becomes negligible after traversing a finite number of synapses. Therefore, the theoretical sum over infinite paths is truncated to a finite depth $K$ for approximation. In this study, $K$ is set to 100, which is sufficient to capture the vast majority of functionally significant pathways. Consequently, the response of a target neuron $n_o$ can be obtained by summing the weights of all paths originating from the LMCs input neuron set with path lengths $k \leq K$. This cumulative weight defines the VFP $M_o$ of the LMCs for each $n_o$ under excitatory and inhibitory contributions.

This truncation approximation converts the problem from summing an infinite series to performing a finite-depth, input-node–initiated constrained graph search[70], achieving computational feasibility while maintaining biological plausibility.

### 3.5 Finite-Depth Connectivity–Based Simulation Model

In the VFP, we obtained representations for all Drosophila brain neurons in the dataset through finite-depth computation, where each $n_o$ corresponds to the VFP $M_o$ of the LMCs under excitatory and inhibitory contributions. In the simulation framework, if a neuron class is blocked, the corresponding neuron nodes are removed during the finite-depth calculation, producing an updated blocked VFP.

The simulation proceeds as follows: for a given stimulus sequence $x$, the Weber contrast $C_W$ is computed from the difference between the current frame $x_t$ and the previous frame $x_{t-1}$, and then fed into the LMCs sensory layer $f_{sence}$ to generate the LMCs input response matrix $r_t$.

This matrix is then dot-multiplied and summed with the VFP $M_0$ of different neurons. After applying low-pass filtering and nonlinear activation, the temporal responses $o_t$ of each neuron at the current time step are obtained. The LP function refers to the low-pass filter.

$$o_t = SiLU\left( LP\left( \sum_{i,j} r_t(i,j)M_0(i,j) \right) \right)$$

It is important to note that, in this model, no additional training or assumed parameters are used beyond the normalization operation.

### 3.6 Neuron Weight Matrix Alignment via Centroid

The centroid of the combined VFP is calculated by summing the matrices across all layers of a neuron (including excitatory and inhibitory contributions) and then computing the centroid, given by:

$$M_o(x_{centroid}, y_{centroid}) = \left( \frac{\Sigma_{i,j}|M_{ij}|\cdot j}{\Sigma_{i,j}|M_{ij}|}, \frac{\Sigma_{i,j}|M_{ij}|\cdot i}{\Sigma_{i,j}|M_{ij}|} \right)$$

Here, $x_{centroid}$ and $y_{centroid}$ indicate the centroid positions along the column and row directions, respectively. The VFP $M_o$ is then shifted so that its centroid aligns with the center, facilitating subsequent analysis and visualization.

### 3.7 ON-OFF Stimulus

In this study, to assess the response properties of visual neurons to changes in luminance, we designed a two-dimensional ON/OFF light block stimulus sequence. Each stimulus image had a size of 41×41 pixels, and the entire grid was divided temporally into two complementary phases: the ON phase and the OFF phase. During the ON phase (from 1000 to 2000 ms), all pixel intensities were set to a high value (1.0) to simulate a sudden increase in brightness. During the OFF phase (from 1000 to 2000 ms), all pixel intensities were set to a low value (0) to simulate a sudden decrease in brightness. The stimulus uniformly covered the entire grid to ensure that neurons received a global luminance change signal. The stimulus sequence was generated with a high temporal resolution of 1 ms to capture the neurons' instantaneous responses.

### 3.8 ON- and OFF-contrast selectivity indices (FRI)

For the ON stimulus sequence $s^{ON}(t)$ and the OFF stimulus sequence $s^{OFF}(t)$, the low-pass filtered neuronal responses $o_{on}(t)$ and $o_{off}(t)$ are computed, and the peak within the analysis window is taken as the representative response for each stimulus type, with negative values set to zero:

$$R_{on} = \max\left(o_{on}(t) - \min\left(o_{on}(t), o_{on}(t)\right), 0\right)$$

$$R_{off} = \max\left(o_{off}(t) - \min\left(o_{on}(t), o_{on}(t)\right), 0\right)$$

The FRI is then calculated as the normalized difference between ON and OFF responses:

$$FRI = \frac{R_{on} - R_{off}}{R_{on} + R_{off}}$$

The FRI ranges from $[-1,1]$, where values close to 1 indicate a preference for bright stimuli (ON). Values close to -1 indicate a preference for dark stimuli (OFF), while values near 0 indicate no clear selectivity between bright and dark stimuli.

### 3.9 DSI and Prefer Angle Stimulus

To assess neuronal direction selectivity for ON-OFF edge motion, we designed two-dimensional dynamic edge motion visual stimuli to mimic the drive of optic flow on direction-selective neurons. Each stimulus image is 41×41 pixels, with edges transitioning from dark-to-light or light-to-dark. The stimuli move along specified directions, covering the full azimuth range (0°–330° in 30° steps). Motion speeds are set at 2, 4, 8, 12, and 16 pixels/ms, with a temporal resolution of 1 ms to capture rapid neuronal responses with high temporal precision. Each stimulus sequence consists of three phases: an initial stationary phase, an edge motion phase, and a final stationary phase. Edge displacement along the motion direction is computed in pixels per frame to ensure full coverage of the stimulus area.

### 3.10 DSI and Prefer Angle

In the simulation, to avoid interference from initial transient responses, the first 500 ms of data are discarded (edge motion begins after 1 s). For each direction $\theta_i$ and speed $v$, the neuronal stimulus sequence $x_{i,v}(t)$ produces a simulated response $o_{i,v}(t)$, and the peak is defined as the maximum of the positive response:

$$R_{i,v} = max(max(o_{i,v}(t), 0))$$

Note that since neuronal responses in the model can approach zero, if the peak variation across directions at the same speed, $\Delta R_v = \max\left(R_v\right) - \min\left(R_v\right) < \delta_{threshold}$, the neuron is considered non-direction-selective.

$$\vec{V} = \sum_v \sum_{i=1}^{N} R_{i,v} e^{j\theta_i}, \quad DSI = \frac{|\vec{V}|}{\sum_v \sum_{i=1}^{N} R_{i,v}}$$

where $j = \sqrt{-1}$, $\theta_i$ is in radians, and $v$ is the edge motion speed. DSI ranges from 0 to 1, with values closer to 1 indicating stronger direction selectivity. The preferred motion direction of the neuron is given by the phase of the vector:

$$\theta_{pref} = arg(\overline{V}), range[0,360]$$

### 3.11 FRI and DSI Mapping Across Neuron Classes

For each neuron class, we first computed the FRI values for all neurons and plotted the median with a 95% confidence interval, retaining only classes with more than 10 samples to ensure statistical reliability. Continuous FRI distributions were generated using Gaussian kernel density estimation[71], normalized, and plotted as symmetric shadows along the X-axis to show distribution shapes, with small random jitter added to display individual data points.

For DSI visualization, we calculated the DSI of each neuron class and generated smooth distributions using kernel density estimation (KDE). To ensure comparability across conditions, ON/OFF responses were time-aligned. The median and 95% confidence interval of DSI were computed using percentiles. For preferred direction visualization, KDE was calculated for each neuron class's preferred angles, plotted on a bi-directional axis: the positive axis for Light→Dark conditions and the negative axis for Dark→Light conditions.

To reflect overall directional preference and strength, weighted average direction and DSI were calculated using a complex vector approach:

$$\overline{DSI}, \bar{\theta} = \sum_{i=1}^{N} DSI_i e^{j\theta_i}$$

where $DSI_i$ is the direction selectivity index of an individual neuron and $\theta_i$ is its preferred direction. The weighted average direction $\bar{\theta}$ and weighted $\overline{DSI}$ reflect the overall directional preference of the neuron class rather than individual neurons.

To intuitively display the relationship between direction preference and DSI, a 2D histogram (angle×DSI) heatmap was plotted in polar coordinates. Angles were binned in 20° increments, and DSI in 0.1 steps. Matrix elements were normalized and color-coded for relative density. Additionally, a weighted average DSI curve was plotted along the angular direction to show overall directional strength:

$$R(\theta) = \arg\left(\sum_{i=1}^{N} DSI_i \cdot 1_{[\theta_i \in bin(\theta)]}\right)$$

The resulting visualization integrates DSI distribution, preferred direction, and directional preference heatmap on a single page, facilitating quantitative comparison across conditions.

### 3.12  Size-Specific Responses of LC Neurons

In this study, we generated moving dark object stimuli of various sizes on a 41×41 pixel canvas, with a gray background (0.5) and dark objects (0). The objects enter from the right side of the canvas and move leftward (corresponding to the front-to-back direction in the fly), with sizes adjustable from 1×1 to 41×41 pixels. The motion speed and frame rate are adjustable, with default values of 20 pixels/sec and 1000 Hz, respectively. Each frame is generated by drawing the object as a square and shifting it horizontally, combined with Gaussian blur to simulate visual edge effects[72,73].

Additionally, grating stimuli consist of alternating bright and dark stripes with a spacing of 5 pixels, moving at 20 pixels/sec. The bright object motion is set identically to the dark object motion, except the object is bright (1). Loom stimuli expand from the center of the canvas, eventually covering the entire 41×41 pixel area.

### 3.13  LC-Based Network

LCNet loads multiple candidate LC types from predefined neuronal VFP files. Each subtype contains VFP for both left and right neurons, recorded in the right-eye visual coordinate system, to construct cross-type visual representations. After loading, the model parses the feature map files and processes each subtype sequentially by name.

During forward computation, the model first reads the LMCs feature vectors corresponding to the left and right eyes. It then loads the VFP of all neuron subtypes and treats them as tensor kernels to compute neuronal responses. For each subtype, the model extracts all right-side VFP and performs a 4D tensor product with the left and right LMCs inputs, generating left and right response scalars for all neurons in that subtype. These responses are concatenated along the last dimension to form the original neuronal response vector for the subtype.

Each subtype is equipped with an independent linear layer that maps its response vector to a fixed 8-dimensional subtype controller representation. The 8-dimensional controllers from all subtypes are stacked and combined via linear weights to produce a fused 8-dimensional visual feature representation. In the final output, task vectors, visual features, and other body-state features are concatenated to form the complete input to the policy network, while the remaining network architecture remains unchanged.

For reinforcement learning, this study follows the framework[3], using Distributed Proximal Policy Optimization (DPPO)[74] as the core training algorithm. DPPO introduces a clipped objective function during policy updates, effectively balancing exploration and stability, ensuring good convergence and robustness in continuous control tasks. During training, the high-level controller optimizes the policy according to a predefined reward function to learn efficient movement and stable navigation strategies in complex terrain. The low-level controller, directly adopted from[3], provides underlying motor patterns and action execution capabilities for the high-level policy without involving high-level control logic.

## 5    Appendix

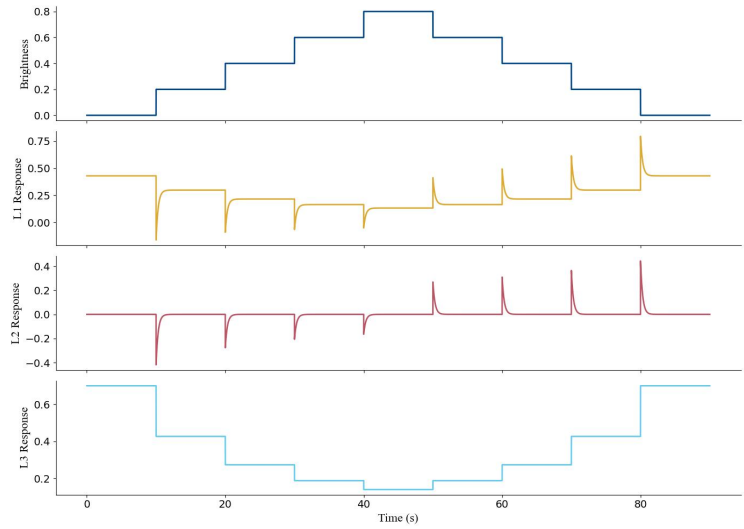

Appendix Figure 1. Response patterns of LMCs under different luminance change stimuli,

fitted based on[39,40].

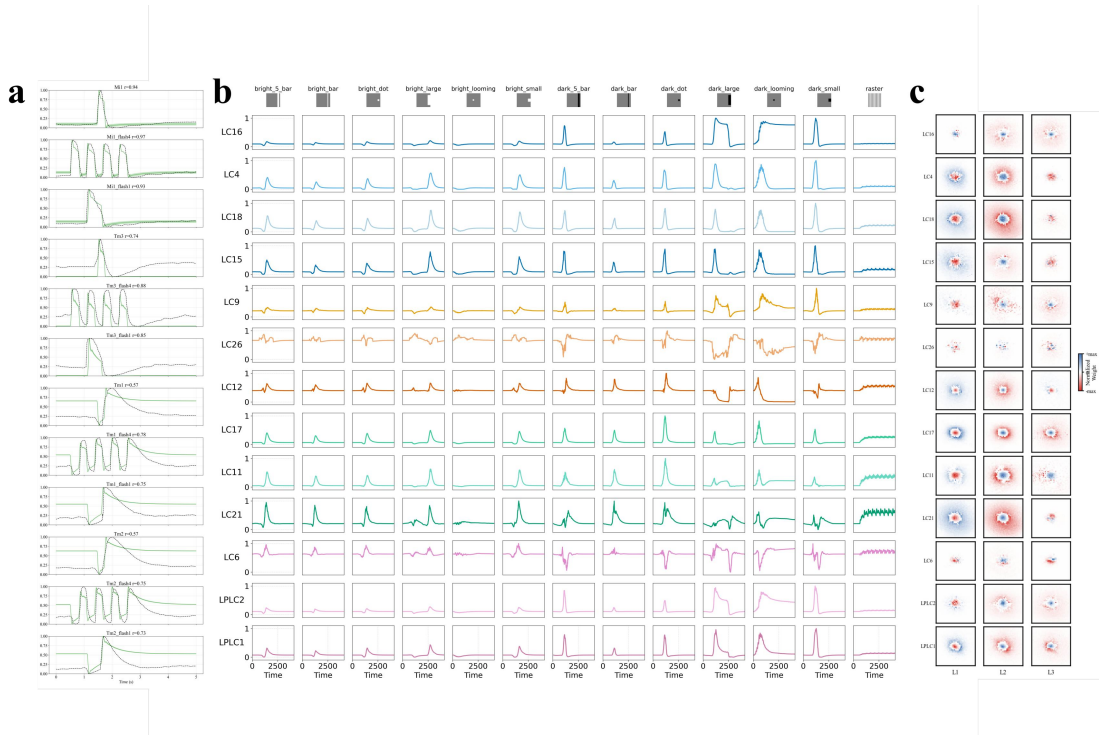

Appendix Figure 2. (a) Comparison of additional physiological experimental data with model

curves and corresponding Pearson correlation coefficients. (b) Median single-neuron response curves across various LC types under different visual stimulus conditions. (c) Average distribution of the shifted VFP for the LC types corresponding to (b).

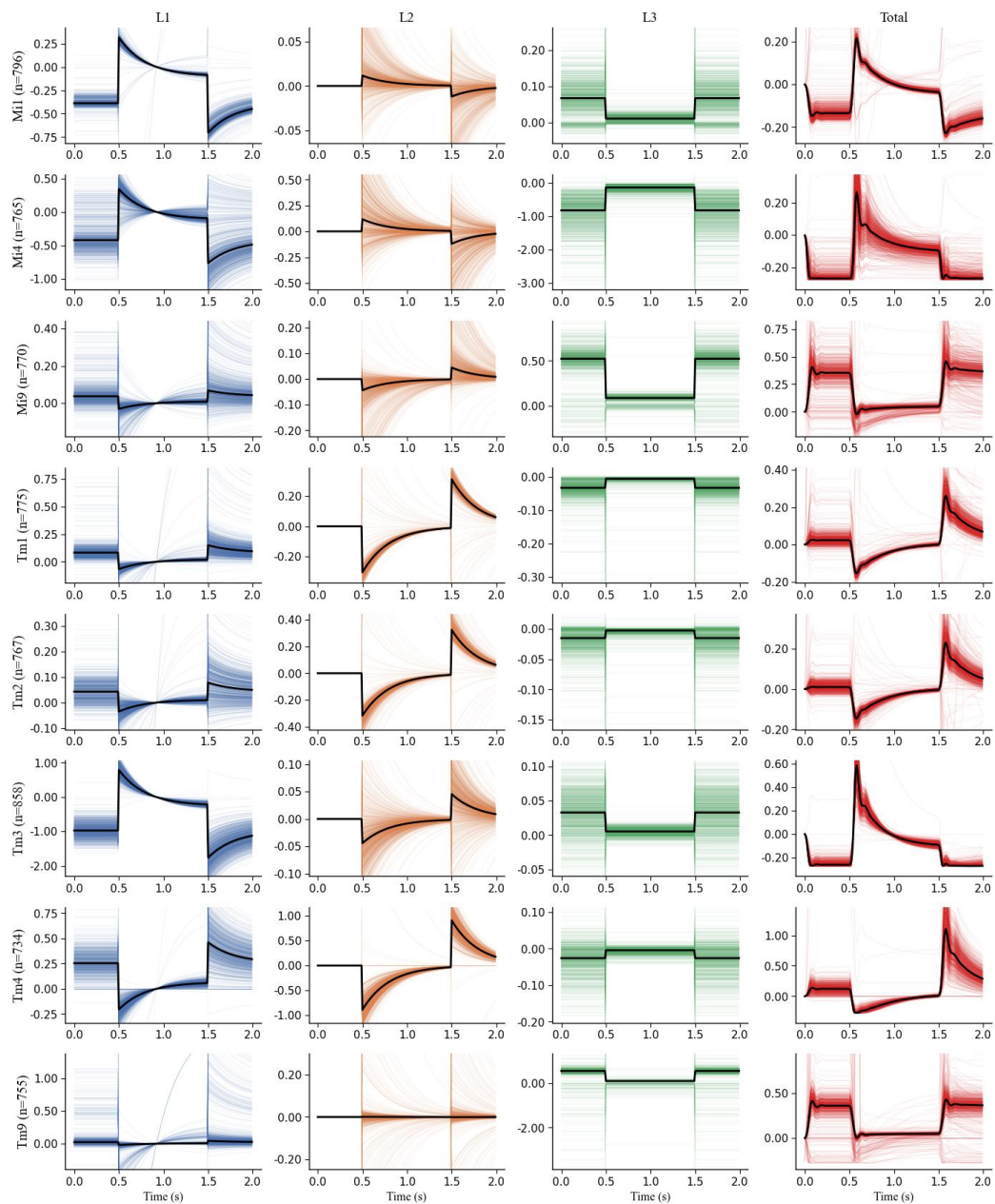

Appendix Figure 3. Responses of selected neuron types (those important for ON/OFF processing, where ON corresponds to 0.5-1.5 s and OFF otherwise) across layers (L1, L2, L3, Total). Black line indicates median, colored faint lines indicate individual neurons.

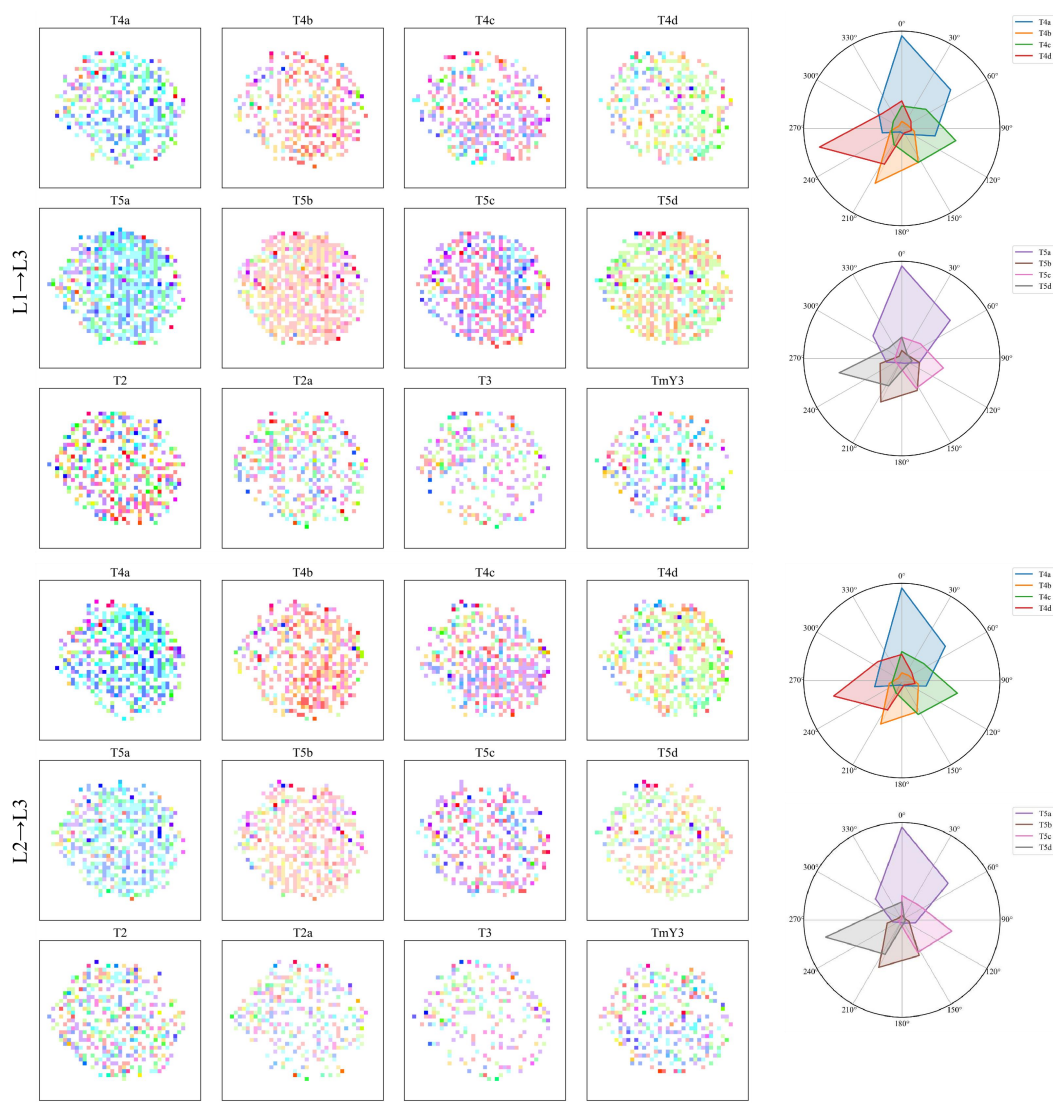

Appendix Figure 4. Spatial vector flow distributions of maximum displacements from L1/L2 to L3 in the VFP for individual neurons of different types, along with their corresponding directional radar plots.

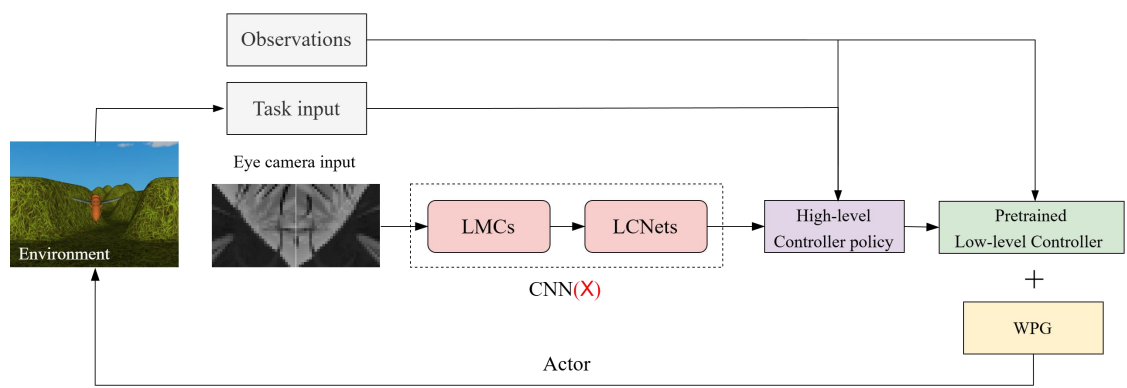

Appendix Figure 5. Steps of reinforcement learning interacting with simulation environments.

**Acknowledgements**

This research was funded by National Natural Science Foundation of China (Grant No.42501093)，and Guangdong Provincial Department of Science and Technology Key Area R&D Program Projects (No. 2022B0101020002). Sponsored by Beijing Nova Program.

**Declarations**

The authors declare no conflicts of interest.

**Data and code availability**

The data and source code are publicly available at https://github.com/jumping321/VFP.git.

**Author Contributions**

**J.P.X.:** conceptualization, methodology, software, writing—original draft preparation. **R.H.R.:** formal analysis, methodology, writing—original draft preparation. **A.Z.:** software, visualization, formal analysis, writing—original draft preparation. **X.Z.:** methodology, formal analysis. **J.S.Z.:** visualization, formal analysis. **W.Y.J.:** conceptualization, methodology, formal analysis, visualization, writing—review and editing, supervision, funding acquisition. **Z.R.Z.:** conceptualization, data curation, formal analysis, writing—original draft preparation.

# Mapping Visual Function Profiles of Individual Neurons Across the Drosophila Brain

## 1. Results of other visual neurons

### a)  FRI

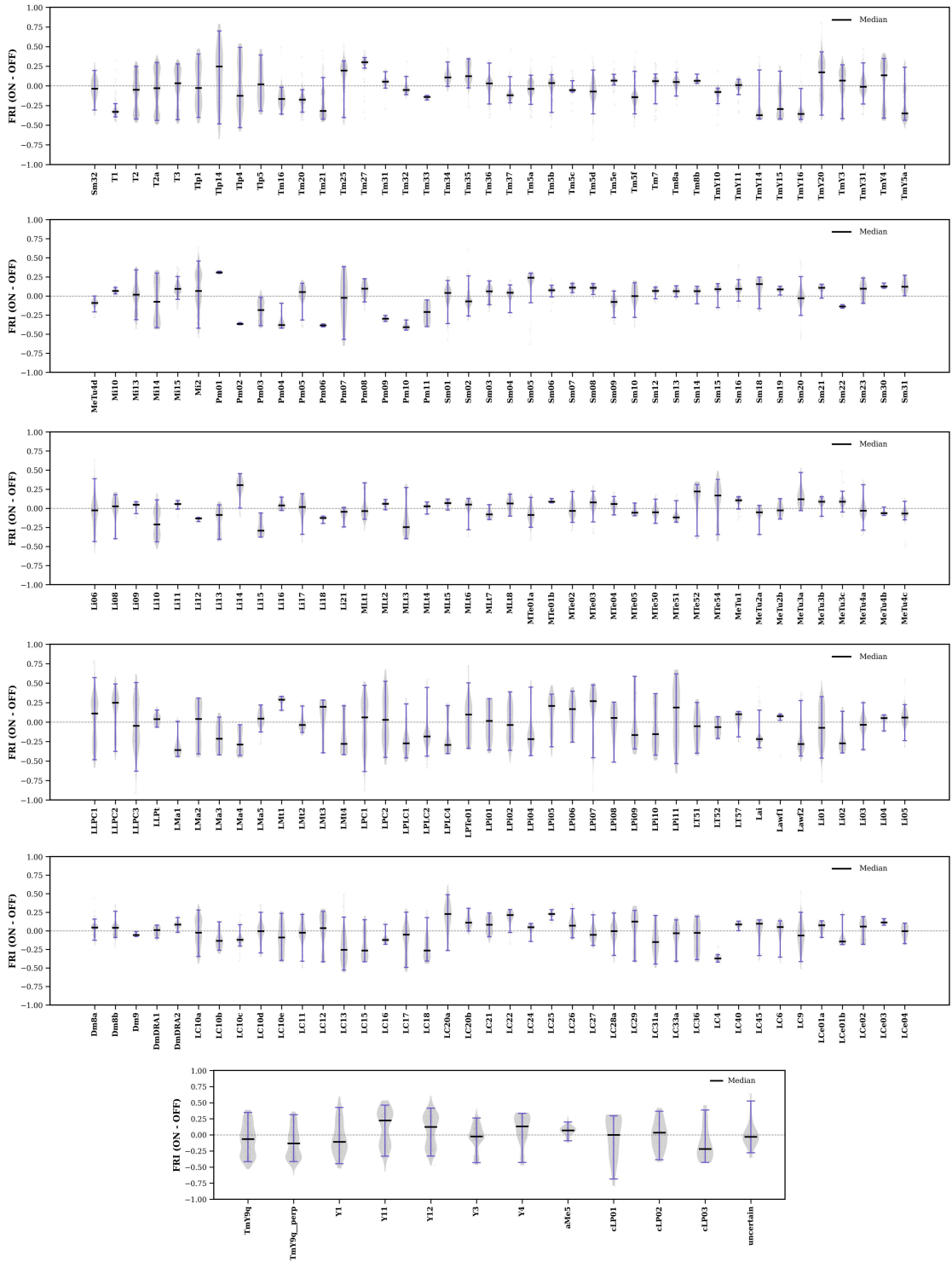

### b)  DSI

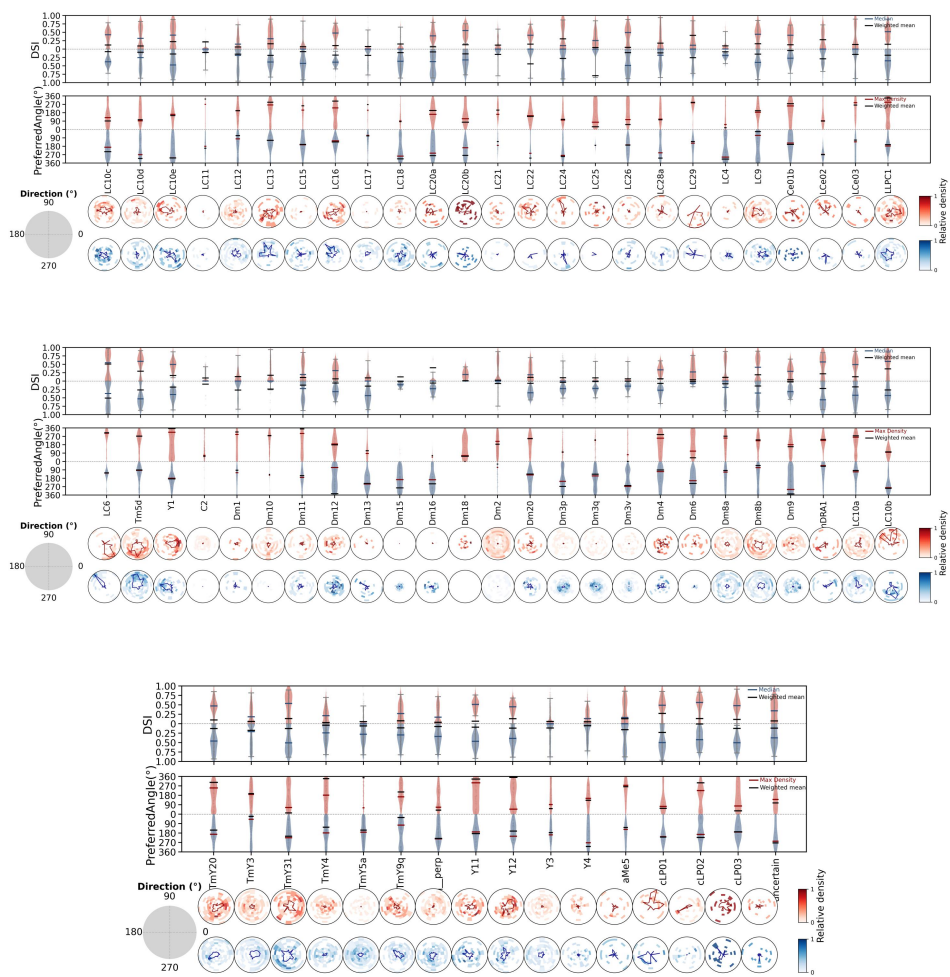

## 2. Visualization

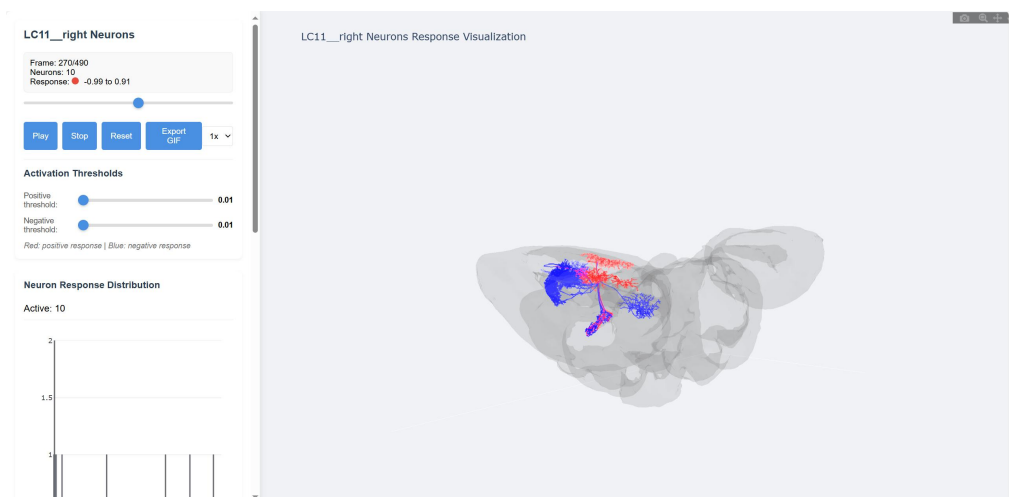

We have provided visualization tools, which will be centrally available in the GitHub repository. The visualization tools utilize the https://navis-org.github.io/navis/ library to illustrate how each neuron in the Drosophila brain responds to different visual stimuli across the visual functional spectrum. Due to the substantial computational resources required for rendering, the visualizations can currently only be displayed in batches.

## 3. Neural Activity data

The data used for Pearson correlation coefficient comparison is stored in neural_activity_data.xlsx(Behnia et al., 2014; Groschner et al., 2022; Silies et al., 2013).

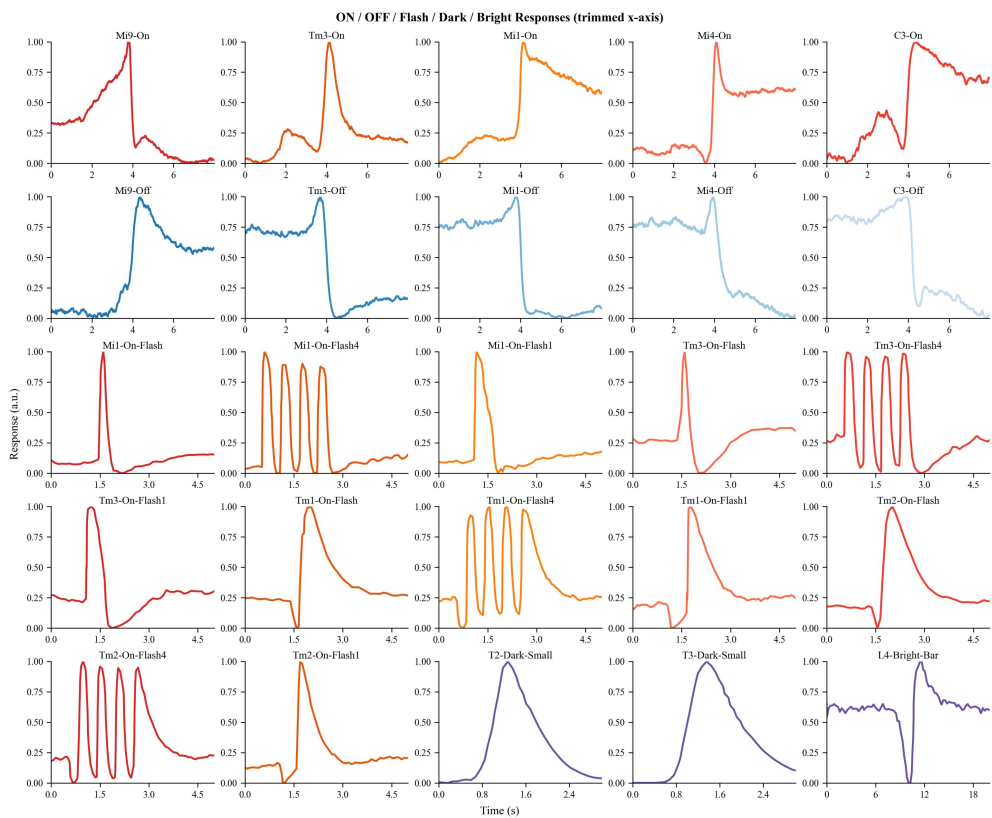

## 4. Code parameter

### a) Multi-Path Aggregation

| target_pos_sum | 1.0 | positive sum |
|---|---|---|
| target_neg_sum | 1.0 | negative sum |
| connections_path | FLYWIRE | connections path |
| synapses_path | FLYWIRE | synapses path |

This code implements a tool for constructing and exporting neural network weight matrices. It reads neuronal connectivity data and synaptic information from two CSV files, then processes them to generate a sparse weight matrix. The tool first computes raw connection weights and then normalizes positive and negative weights separately per column toward specified target sums. The entire computation process is parallelized for efficiency and employs chunked processing for memory management. The final output is saved in three formats: an efficient binary format for sparse matrices, a list of neuron identifiers, and a human-readable text format suitable for further analysis, visualization, or as input to other computational tasks.

| neuron_types | 'L1','L2','L3' | neuron types to analyze |
|---|---|---|

| side | 'right' | side to analyze |
|------|---------|-----------------|
| max_depth | 100 | maximum search depth |
| min_weight | 1e-8 | minimum weight threshold |
| blocked_types | ['T3'] | blocked neuron types |

Multi-Path Aggregation method implements a neural circuit analysis pipeline that leverages connection data to compute weighted path relationships between specific neuron classes. It begins by loading a precomputed sparse weight matrix and corresponding neuron identifiers, then integrates cell type annotations to classify neuronal populations. The analyzer performs parallelized graph traversal to identify valid synaptic pathways up to a defined depth, filtering connections below a specified weight threshold. During computation, certain neuron types can be excluded to model specific circuit conditions or experimental manipulations. Results are systematically saved for downstream applications such as connectivity mapping, circuit motif analysis, and network modeling. The modular design allows researchers to flexibly investigate information flow through defined neural subpopulations under configurable constraints.

**b)  LMCs parameter**

| Type | Function Form | Description |
|------|---------------|-------------|
| L1 Static Response | 0.35×exp(-2.36x)+0.08 | L1 layer neuron static response curve |
| L3 Static Response | 0.62×exp(-2.90x)+0.08 | L3 layer neuron static response curve |
| L1_B | -0.917·x/(1+|1.992·x|) | L1 layer B dynamic response |
| L1_D | -2.326·x/(1+|-5.377·x|) | L1 layer D dynamic response |
| L2_B | -0.814·x/(1+|1.950·x|) | L2 layer B dynamic response |
| L2_D | -2.044·x/(1+|-3.623·x|) | L2 layer D dynamic response |
| L1_DECAY_TAU | 300 | Decay time constant for L1 layer neurons |
| L2_DECAY_TAU | 300 | Decay time constant for L2 layer neurons |
| L3_DECAY_TAU | 200 | Decay time constant for L3 layer neurons |

**B (Brightening) and D (Darkening) responses** (Ketkar et al., 2022; Ketkar et al., 2020) represent the two opposing response modes of the same lamina monopolar neurons to contrast changes. Each neuron exhibits dual-state encoding: B-mode describes its response to luminance increments (light-ON), characterized by positive saturation dynamics, while D-mode captures its response to luminance decrements (light-OFF), showing negative saturation behavior. This bidirectional coding within individual neurons enables efficient contrast processing through a single cellular pathway, where the same neural element differentially processes brightening and darkening signals based on input polarity. The distinct parameter sets (positive b-values for B, negative for D) mathematically define how each neuron separates and encodes opposing contrast directions through a unified sigmoidal response framework.

### c) Model parameter

| Parameter Name | Value/Type | Description |
|---|---|---|
| NEURON_GRID | (41, 41) | Neuron grid size |
| TIME_STEP | 1 | Time step |
| LOW_PASS_CUTOFF | 10 | Low-pass filter cutoff frequency |
| SAMPLING_RATE | 1000 | Sampling rate |
| Butterworth order | 4 | Low-pass filter order |
| Filter type | 'low' | Butterworth low-pass |

This model implements an object size response simulation system through a three-layer neural circuit architecture (L1, L2, L3) with distinct temporal dynamics and response characteristics. It processes visual stimuli through layer-specific computations: L1 and L3 layers combine both static and dynamic response components, while L2 focuses exclusively on dynamic contrast encoding. The system loads pre-defined spatial weight matrices representing neuronal receptive fields, applies Weber contrast calculations to determine brightening (B-type) and darkening (D-type) responses, and integrates these signals through temporal filtering and SiLU activation. The implementation achieves biologically plausible simulations through numerically stable operations including global weight normalization, proper contrast handling, and exponential decay dynamics.

### d) LC-Based Model parameter

The LMCsProcessor class in the 'task.py' code implements a Lamina Monopolar Cells (LMCs) visual preprocessing model, which is biologically inspired by the early visual processing stages in insect compound eyes. It simulates three types of LMC neurons (L1, L2, L3) that process raw visual input from photoreceptors.

Key Additional Input Variables:

1.  time_step (int, default=1)

2.  tau_l1 (int, default=9)

3.  tau_l2 (int, default=9)

4.  tau_l3 (int, default=6)